\documentclass[%
 reprint,
 pra
]{revtex4-2}

\usepackage{graphicx}
\usepackage{dcolumn}
\usepackage{bm}
\usepackage{url}
\usepackage{placeins}

\begin{document}

\preprint{APS/123-QED}

\title{Dynamic compensation of stray electric fields in an ion trap using machine learning and adaptive algorithm}

\author{Moji Ghadimi$^{1}$}
 \email{m.ghadimi@griffith.edu.au}
\author{Alexander Zappacosta$^{1}$}
\author{Jordan Scarabel$^{1}$}
\author{Kenji Shimizu$^{1}$}
\author{Erik W Streed$^{1, 2}$} 
\author{Mirko Lobino$^{1, 3}$}

\affiliation{%
$^{1}$Center for Quantum Dynamics, Griffith University, Nathan QLD Australia \\ 
$^{2}$Institute for Glycomics, Griffith University, Southport QLD  Australia \\
$^{3}$Queensland Micro Nanotechnology Centre, Nathan QLD Australia
}%

\date{\today}

\begin{abstract}
Surface ion traps are among the most promising technologies for scaling up quantum computing machines, but their complicated multi-electrode geometry can make some tasks, including compensation for stray electric fields, challenging both at the level of modeling and of practical implementation. Here we demonstrate the compensation of stray electric fields using a gradient descent algorithm and a machine learning technique, which trained a deep learning network. We show automated dynamical compensation tested against induced electric charging from UV laser light hitting the chip trap surface. The results show improvement in compensation using gradient descent and the machine learner over manual compensation. This improvement is inferred from an increase of the fluorescence rate of 78\% and 96\% respectively, for a trapped $^{171}$Yb$^+$ ion driven by a laser tuned to -7.8 MHz of the $^2$S$_{1/2}\leftrightarrow^2$P$_{1/2}$ Doppler cooling transition at 369.5 nm.
\end{abstract}

\maketitle


\section{Introduction}

Trapped ions are one of the most promising candidates for implementing scalable quantum information processing \cite{debnath_demonstration_2016, lekitsch_blueprint_2017,monroe_scaling_2013, blinov_quantum_2004}. Traditionally macroscopic rod or needle traps were used for electrically trapping laser cooled ions. These Paul traps have a deep potential well that enables the trapping of multiple ions in a chain, but their distant electrodes lack the fine spatial control of electric fields necessary to efficiently split and combine small ion crystals \cite{wineland_experimental_1998}. One of the proposed architectures for scaling up the number of trapped ions in a quantum processor is the use of microfabricated surface chip traps with multiple DC electrodes that are able to manipulate multiple ions individually \cite{kielpinski_architecture_2002, chiaverini_surface-electrode_2005, seidelin_microfabricated_2006, doret_controlling_2012, amini_toward_2010}. 

In these chip traps the ions are confined to a potential formed by the node of an oscillating RF electric field in two dimensions while a DC field generated from multiple electrodes provide a finely spatially adjustable potential in the third dimension. It is well-known in Paul trapping that stray DC electric fields push the ion off the RF node and induce micro-motion that degrades or outright prevents effective confinement and laser-cooling \cite{kielpinski_entanglement_2001}. The solution to this problem is to use the DC electrodes to generate an opposite DC electric field that compensates for these stray fields \cite{doret_controlling_2012}. 
 
Micro-motion amplitude can be measured and dealt with in multiple ways \cite{tanaka_micromotion_2012, allcock_heating_2012, narayanan_electric_2011, doret_controlling_2012, berkeland_minimization_1998}. A simple and quick proxy for small amounts of micromotion is to measure the ion’s fluorescence under laser cooling near the Doppler limit. In fact for low magnitudes of micro-motion, if a laser is red-detuned near the natural linewidth of the atom, the ion fluorescence rate increases as the micro-motion amplitude decreases (see Section \ref{sec:exp} and Fig. \ref{fig:microm}). While stray electric fields can be readily compensated in simple fixed trap geometries \cite{noek_trapping_2013}, this task becomes more challenging for multi-electrode designs \cite{allcock_implementation_2010}  and in proximity of surfaces which are vulnerable to laser induced charging \cite{harlander_trapped-ion_2010}. 

Automation, optimization, and machine learning have been used to improve different manual tasks in atom and ion traps \cite{seif_machine_2018, wigley_fast_2016, tranter_multiparameter_2018} and they are also useful tools for optimizing the individual electrode voltages which generate the trapping electric field. In this work we show how the voltages of an array of electrodes in a surface chip trap can be optimized using a fully automated, machine learning driven process to minimize micromotion and maximize florescence. The chip trap in our experiment has 44 segmented DC electrodes and 2 RF rails, as shown in Fig.~\ref{system_diagram}. The trap incorporates nearly aberration free diffractive mirrors directly fabricated onto the central ground electrode for efficient fluorescence collection from the ion \cite{ghadimi_scalable_2017}. A Gradient descent (ADAM \cite{kingma_ADAM_2017}) and a machine learning algorithm (MLOOP \cite{wigley_fast_2016}) were compared on the basis of the versatility and time taken to find the optimal compensation and highest fluorescence rate of the ion. These methods were applied to a trapped $^{171}$Yb$^+$ ion and compared to a manual optimization performed by an experienced operator. 

With ADAM we were able to improve the fluorescence by 78\% starting from voltages already optimized by manual adjustment, while MLOOP achieved a 96\% improvement. Subsequently we tested the versatility and adaption of this procedure by deliberately charging the trap with UV light to drop the fluorescence rate by around 35\% and compensating back to the optimal with ADAM. This was not tested on MLOOP as it was observed to be quite sensitive to the noise in the fluorescence signal and did not make a solid candidate for final optimization of the artificially charged trap.

\section{Experimental Setup}
\label{sec:exp}

\begin{figure*}
 
 \includegraphics{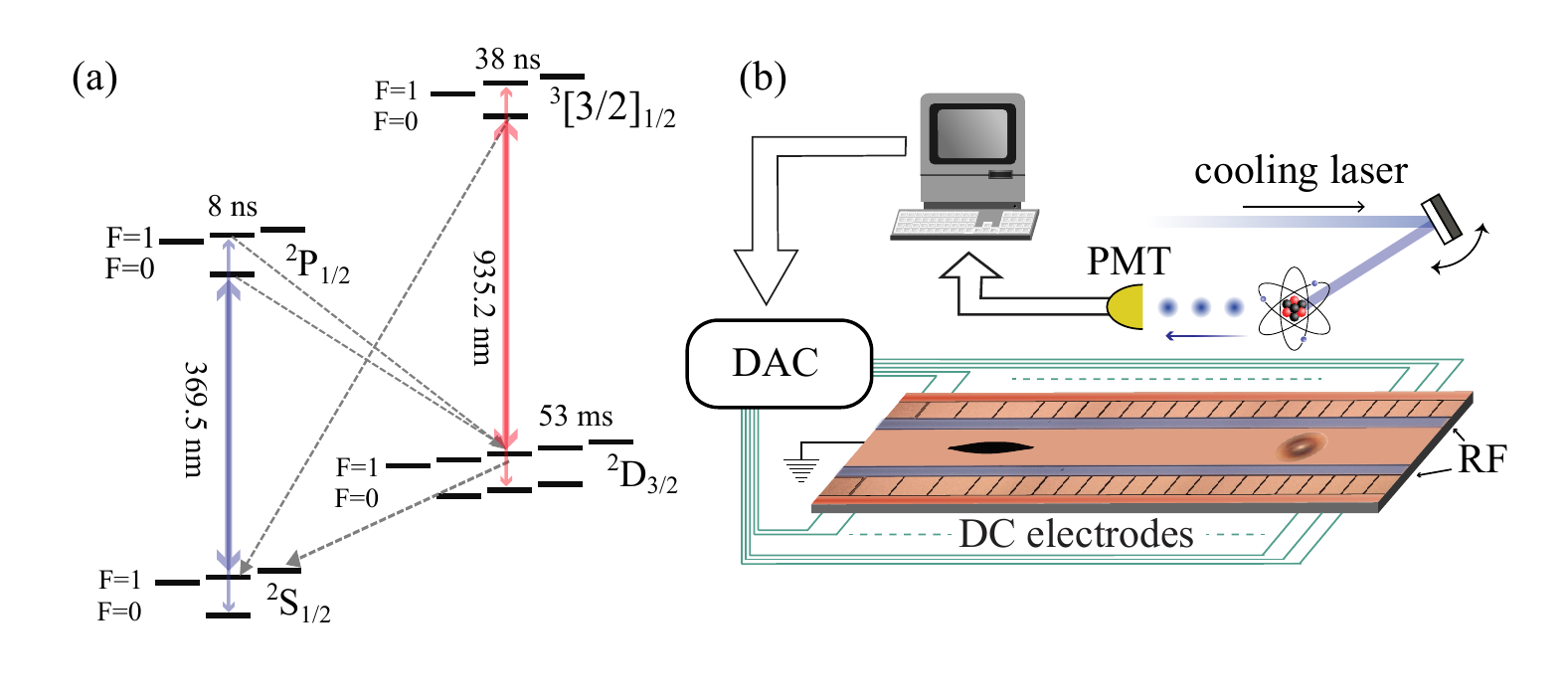}
 \caption{a: Transitions for laser cooling $^{171}$Yb$^+$. The ion is laser cooled on the nearly closed 396.5~nm
$^2$S$_{1/2}$F=1 to $^2$P$_{1/2}$F=0 transition. The small amount of off-resonant scattering into the $^2$S$_{1/2}$ F=0 state is repumped with a 14.7~GHz sideband on the primary cooling laser. Decays from $^2$P$_{1/2}$ can also populate the $^2$D$_{3/2}$ level, which are repumped using a 935~nm laser. b: Schematics of the ion trap chip and the optimization devices. RF and DC electrodes are used to trap the ion and shuttle it above the diffracting mirror that collects photons from the ion and sends it to photo multiplayer tube (PMT). The PMT counts are processed by the optimiser and DC electrode voltages and cooling laser position are updated using a digital to analogue converter (DAC).}
 \label{system_diagram}
\end{figure*}

The ion trap used in these experiments and described in \cite{shappert_spatially_2013,ghadimi_scalable_2017},  consists of a planar rectangular chip of 1400~$\mu$m length and 800~$\mu$m width. It has a set of aluminum electrodes patterned on top and is schematically shown in Fig.~\ref{system_diagram}. The ion is trapped 60 microns above the surface, between the two RF electrode rails that extend for the full length of chip, and 44 DC electrodes on the sides. The DC electrodes are used to create a reconfigurable trapping potential along the length of the trap, compensating stray electric fields, and enabling ion crystals merging, separation, and shuttling. 

An atomic oven underneath one end of the trap generates a beam of neutral Yb atoms which passes through a slit in the chip. The $^{171}$Yb is first excited by an isotopically selective 399~nm laser \cite{blums_laser_2020}, and subsequently non-resonantly ionized by the a 369.5~nm laser. The ion is then Doppler cooled using the same 369.5~nm laser tuned nearly resonant with $^2$S$_{1/2}$~F=1$\leftrightarrow^2$P$_{1/2}$~F=0 transition. Occasionally off-resonant scattering from the $^2$P$_{1/2}$~F=1 state will populate the dark $^2$S$_{1/2}$~F=0 ground state, which is repumped by a small amount of light from a 14.7 GHz sideband added to the primary 369.5~nm cooling laser. There is also a probability of 0.5\% for the atom to decay from the $^2$P$_{1/2}$ into the meta-stable $^2$D$_{3/2}$ state. In this case a 935~nm laser repumps the ions back into the cooling cycle. Figure~\ref{system_diagram} depicts the cooling and repumping transitions.

After initial trapping near the oven slit region, the ion is shuttled along the length of the trap by properly controlling the voltages of the array of DC electrodes with a 12-bit National Instruments PXI-6713 DAQ (output doubled to +/-20V with 10 mV resolution). For our experiments, the ion was shuttled above the surface of an integrated diffracting micro-mirror patterned on the ground electrode with a focal length of 60~µm, equal to the height of the ion above the trap, to improve florescence collection and coupling into a single mode fiber \cite{ghadimi_scalable_2017}.

\begin{figure}[b]
\includegraphics{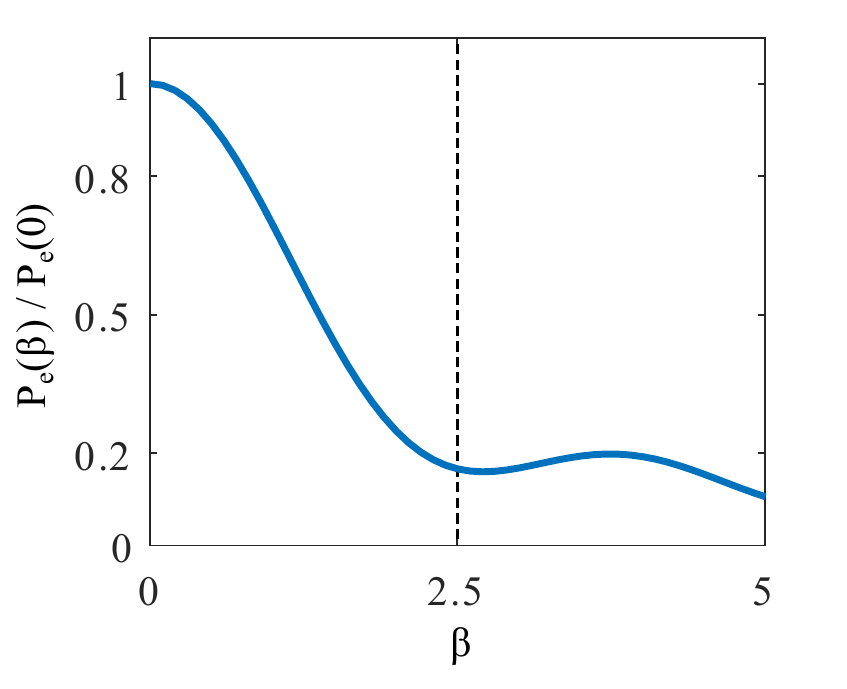}

 \caption{Normalised florescence as a function of the dimensionless micromotion parameter $\beta$ (Eq. \ref{equ:microm}). Here $\delta_L$, the laser frequency detuning, is -7.8 MHz, $\gamma$, the natural linewidth is 20 MHz and $\Omega$, the RF trap frequency, is 25.5 MHz. The $^2$P$_{1/2}$ state population and therefore the fluorescence rate drops monotonically as $\beta$ increases as long as its value is not too large (smaller than 2.5). Our operating regime is far below this threshold.}
 \label{fig:microm}
\end{figure}

For efficient Doppler cooling and subsequent quantum information processing experiments, it is important to minimize any micro-motion induced by stray electric fields. This is done by tuning the voltages of the DC electrodes so that they generate an opposing electric field that compensates for the stray fields. To generate a compensation field in the RF confinement plane $xy$, we use two waveforms (array of DC voltages for individual electrodes) which were calculated for generating an electric field of 100 V/m in the $x$ (horizontal) and $y$ (vertical) directions. Electrode voltages of $-1$ to $+1$~V for the $x$ direction and $-2$ to $+2$~V for the $y$ direction are needed to create electric field of 100 V/m at the location of the ion. These sets of voltages are multiplied by arbitrary weights $w_x$ and $w_y$ to compensate arbitrary fields in the RF trapping plane. In practice this process has imperfections and cannot fully compensate for the stray electric fields \cite{allcock_implementation_2010}. Instead, individual electrodes need to be tuned independently for optimal compensation. This optimum compensation is position dependent and there might be multiple solutions for a multi-electrode trap.

Two other DC waveforms are used and their relative weights tuned: a harmonic waveform, and RF plane trap axis rotation waveform. The former creates a harmonic potential to trap the ion in different locations along the chip length, while the latter rotates the trap's secular confinement axes to lift the degeneracy and improve cooling \cite{norton_millikelvin_2011}. 

We infer the magnitude of the stray electric field from the fluorescence rate of the ion when the cooling laser is red-detuned near half its natural linewidth for optimal cooling. When driving an ideal cold two-level atom with this laser, the fluorescence is proportional to the population of excited state, $P_e$, that can be calculated from the optical Bloch equation \cite{budker_atomic_2004}. The relationship between $P_e$ and the strength of the stray electric field is given by \cite{berkeland_minimization_1998}:   

\begin{equation}
       P_e(\beta) = C \sum_{m=-\infty}^{\infty}{\frac{J_m^2(\beta)}{ (\frac{\delta_L}{\gamma}+m\frac{\Omega}{\gamma})^2 + \frac{1}{4}   }}
       \label{equ:microm}
\end{equation}

where $J_m$ is the m\textsubscript{th} order Bessel function of the first kind, $\delta_L=\omega_{atom}-\omega_{laser}$ is the laser frequency detuning, $\gamma$ is the natural linewidth and $\Omega$ is the RF trap frequency. $\beta$ is a dimensionless measure of the ion's coherently driven motion from the stray electric fields and RF phase imbalances (AC micromotion) \cite{berkeland_minimization_1998}. $C$ depends on the strength of the coupling between the levels that is a constant here since we keep laser intensity and direction fixed. Figure \ref{fig:microm} shows a graph of $P_e(\beta)/P_e(0)$ as a function of $\beta$. The graph shows that at our detuning (-7.8 MHz) the population decreases monotonically as $\beta$ increases if $\beta$ is less than 2.5. For larger $\beta$ micromotion induced local maxima arise that prevents inferring the magnitude of micromotion from fluorescence rate. Since our operating regime is below this threshold, we can use the change in fluorescence rate to detect the change in magnitude of the stray electric field.

\section{Results}

A gradient descent algorithm (ADAM) and a deep learning network (MLOOP) were tested for compensating stray fields in different working regimes. The source code used for the experiments is available in \cite{github_Zapp}. The software controlled the voltages using the PXI-6713 DAQ and read the fluorescence counts from a photo-mutliplier-tube (PMT) through a time tagging counter (IDQ id800). All software was written in python and interfaced with the DAQ hardware using the library NI-DAQmx Python. A total of 44 DC electrodes and the horizontal position of the cooling laser were tuned by the program, resulting in a total of 45 input parameters.  

\subsection{Gradient Descent optimizer }
\label{ADAM}

The first compensation test was performed by ADAM gradient descent algorithm. This is a first order optimizer that uses the biased first and second order moments of the gradient to update the inputs of an objective function, and was chosen for its fast convergence, versatility in multiple dimensions and tolerance to noise\cite{kingma_ADAM_2017}. Our goal was to maximize the fluorescence of the ion which was described by a function $f(\vec{\alpha}{\,})$, where $\vec{\alpha}{\,} =(\alpha_1, \alpha_2, \alpha_3, …\alpha_{45})$ represents the array of parameters to be optimized. To find the optimal $\vec{\alpha}{\,}$, the algorithm needs to know the values of the partial derivatives for all input parameters. Because we do not have an analytic expression for $f(\vec{\alpha}{\,})$, the values of its derivatives were estimated from experimental measurements by sequentially changing each input $\alpha_{i}$, and reading the associated change in fluorescence $f$. This data were used as inputs to ADAM for finding the optimal $\vec{\alpha}{\,}$ which maximized $f$.

\begin{figure}[b]
\includegraphics{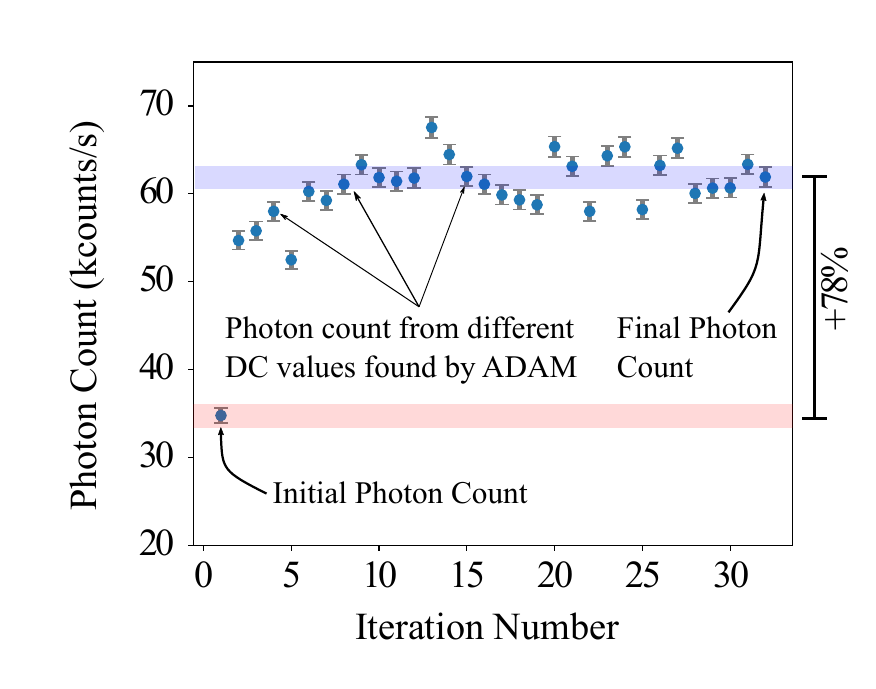}
 \caption{The photon count improvement using ADAM starting from manually optimized waveform weights. This graph demonstrates a  $78\pm1\%$ improvement in ion fluorescence. The  error bars are a combination of the readout error of the PMT and peak to peak variations in the photon count whilst not optimizing. The background photon count here is 1184$\pm$34~counts/s. }
 \label{ADAMgraph}
\end{figure}

Before running the automated compensation, we manually adjusted the 4 weights of the waveforms used for compensation described in the previous section. We also tried to run ADAM to optimize these 4 parameters but the increase in fluorescence was limited to 6\%. 
After manual compensation, we ran ADAM on all 45 inputs with the algorithm parameters given in the source code \cite{github_Zapp}. Each iteration took 12~s, where 9.8~s were the photon readout (0.1~s×2 readouts per parameter plus 2×0.1~s readouts at the beginning and end of the iteration), and the rest of the time was the gradient computation. If the photon count reduced by more than 40\% of its initial value, the algorithm terminated and applied the previously found optimum. This acted as the safety net for the program, ensuring the ion was not lost while optimizing the 45 inputs. We need this safety net because if the ion is heated past the capture range for the used cooling detuning, it will be ejected from the trap. 

\begin{figure*}
 \includegraphics{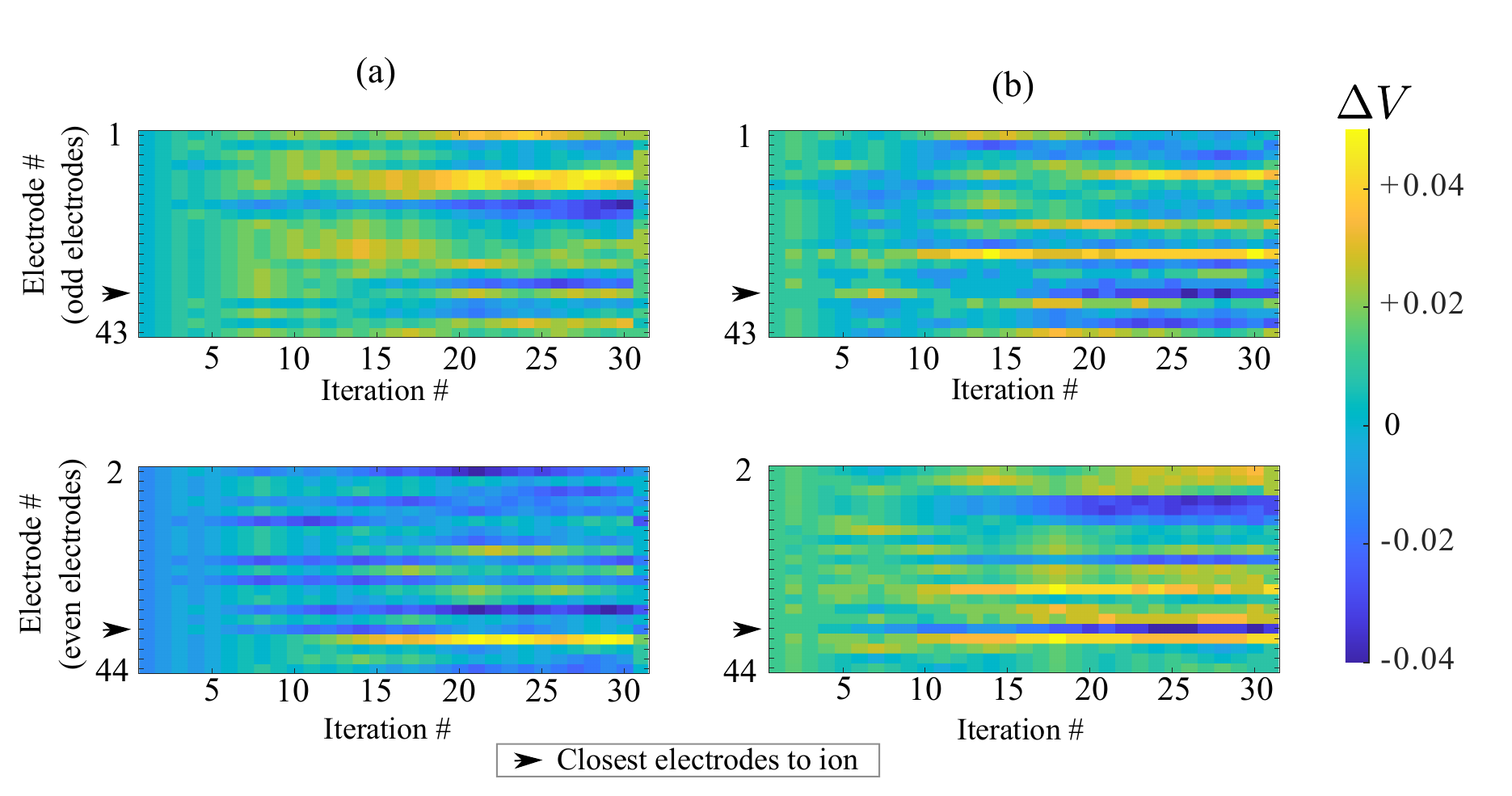}
 \caption{Voltage deviation from the original starting point during optimization with ADAM. a: uncharged trap (section~\ref{ADAM}). b: During UV charging (section~\ref{poor_con}). Top graphs show odd electrode numbers corresponding to top DC electrodes in Fig. \ref{system_diagram} and the bottom graphs show the even electrode numbers. The values were determined by subtracting the voltage at each iteration by the starting voltage $\Delta V = V_n - V_0$. Changes can be seen in almost all the electrodes of the trap. }
 \label{ChangeInVoltage}
\end{figure*}

In our implementation of the algorithm we removed the reduction in the step size of the optimization algorithm as iterations progressed. This step reduction, which is present in the standard version of ADAM, is not ideal when stray fields change with time since the optimal values of the voltages also drift in time. The removal caused some fluctuations in the photon readout near the optimal settings. Adding to these fluctuations, other sources of noise, such as wavemeter laser locking \cite{ghadimi_multichannel_2020}, and mechanical drift in the trap environment, resulted in daily photon count variations of around 5\%. Despite these fluctuations, and the fact that stray fields change every day, the algorithm demonstrated an increase in fluorescence collection up to 78±1\% (Fig.~\ref{ADAMgraph}) when starting from a manually optimized configuration.

The ADAM algorithm was fast and reliable (the ion was never lost during optimization), even in extremely volatile conditions like having time-dependent charging and stray electric field buildup. Figure~\ref{ChangeInVoltage}a shows a colourmap of the voltages and laser position adjustments, where most of the improvement came from adding the same voltage to all DC electrodes indicating the ion was not at optimal height. The volatility of the ion-trap environment causes the fluorescence rate to oscillate around the optimal point. To get the best value, instead of using the values of the final iteration, the software saved all voltage combinations and applied the setting with the highest photon count after all iterations were finished. Despite picking the best value it can be seen in Fig.~\ref{ADAMgraph} that the fluorescence for some iterations during the optimization are higher than the final point selected by the software. This is because when the settings are changed, the ion fluorescence rate may transiently increase and subsequently stabilize to a slightly lower value for the same voltage settings.

\subsection{Deep Learning Network for the Ion trap}

\begin{figure}
 \includegraphics{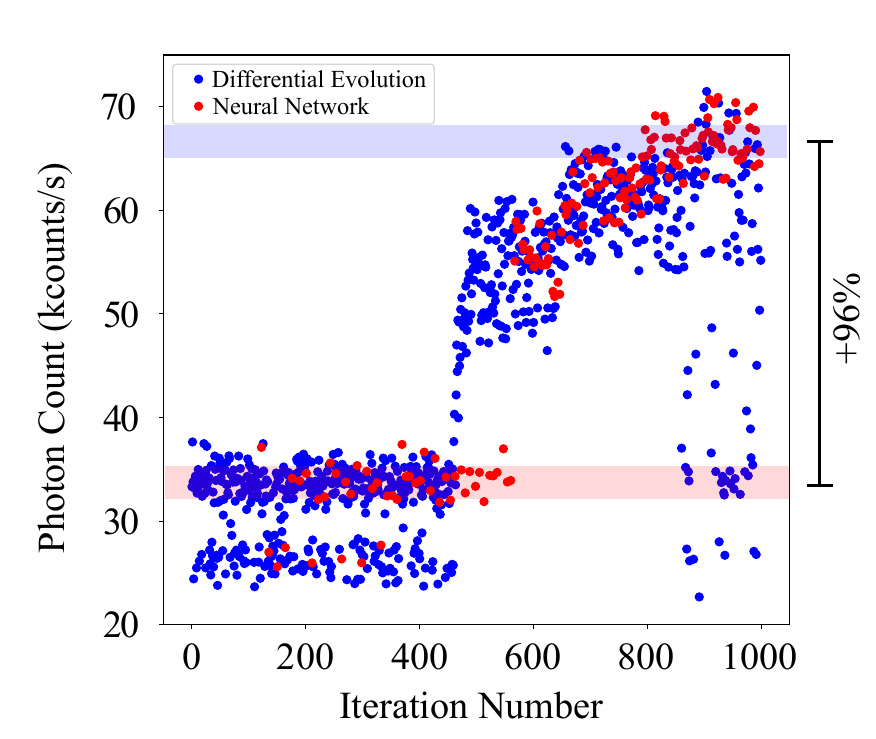}
 \caption{MLOOP deep learning network. Differential Evolution explores the input space (blue points) and the neural network creates a model of the data and predicts an optimum (red points). Maximum photon count of the neural network points is 96$\pm$1\% higher than manual optimization. Differential Evolution continues to explore the input space and has varied photon counts. The beginning point for the process (found by manually adjusting the 4 waveform weights) was at 33700~counts/s and the highest photon count found by the neural network was at 66200~counts/s.}
  \label{MLOOP}
\end{figure}

The second algorithm tested was a deep learning network using the python based optimization and experimental control package MLOOP\cite{wigley_fast_2016}. MLOOP uses Differential Evolution \cite{price_differential_2005} for exploring and sampling data. The blue points in Fig. \ref{MLOOP} corresponds to these samples and it can be seen that even at the end of optimization, they can have non-optimum fluorescence rates.

MLOOP also trains a neural network using the data collected by Differential Evolution and creates an approximate model of the experimental system. It then uses this model to predict an optimum point. The red points in Fig. \ref{MLOOP} shows the optimum points predicted by the neural network model. It can be seen that this section starts later than Differential Evolution, as it requires some data for initial neural network training, and gradually finds the optimum and stays near it. For training of the neural network, the inbuilt ADAM optimizer is used to minimize the cost function. 

The sampling in MLOOP does not require a gradient calculation which greatly improves the sampling time. Even though the sampling is fast, training the network to find an optimal point requires a minimum of 100 samples and that makes MLOOP slower than ADAM. With our settings for MLOOP, each iteration took 0.7~s on average and therefore 700~s was needed to take 1000 samples shown in Fig. \ref{MLOOP}.  

In our test the neural network in MLOOP had 5 layers with 45 nodes each, all with Gaussian error correction. The neural network structure was manually optimized and tested on a 45-dimensional positive definite quadratic function before being used for the experiment. 

Once the ion was trapped, positioned above the integrated mirror \cite{ghadimi_scalable_2017}, and photon counts were read, the program started sampling 100 different voltage combinations around its initial point. Then, the network started training on the initial data and making predictions for the voltages that maximise fluorescence. 

Since the ion trap setup is very sensitive to changes in the electric field, the voltages were set to move a maximum of 1\% of their previous value in each iteration to reduce the chance of losing the ion. As a step size value could not be explicitly defined, this percentage was chosen to make the changes similar to the step size used for ADAM. 

A small percentage of our initial trials with the maximum change of a few percent (instead of 1\%) led to an unstable ion during the parameter search sequence. This is because MLOOP is a global optimizer and can set the voltages to values far from the stable starting point. Since the ion trap is a complicated system that can only be modelled for a specific range of configurations, moving away from these settings can lead to unpredictable and usually unstable behavior.  

MLOOP also has an in-built mechanism that handles function noise using a predefined expected uncertainty. We set this uncertainty to the peak-to-peak noise of the photon readout when no optimization was running. 

Since MLOOP is a global optimizer it was able to find optimum points different from the points found by ADAM. For trials where low numbers of initial training data points were used, these configurations proved to be unstable and in most cases resulted in the loss of the ion. Unstable states were also observed occasionally if the optimizer was run for too long. With moderate-size training sets, MLOOP was able to find voltage settings with fluorescence rates similar or higher than optimum points found by ADAM as shown in Fig.~\ref{MLOOP}.   

Considering the long duration of the MLOOP iteration sequence and the possibility of finding unstable settings in volatile conditions, the test of optimization with induced changing stray fields (section \ref{poor_con}) was only performed with the ADAM optimizer as the gradient based search method proved to be more robust against fluctuations in the ion environment.

\subsection{Ion Properties before and after optimization}
\begin{figure}[b]
 \includegraphics[width=0.5\textwidth]{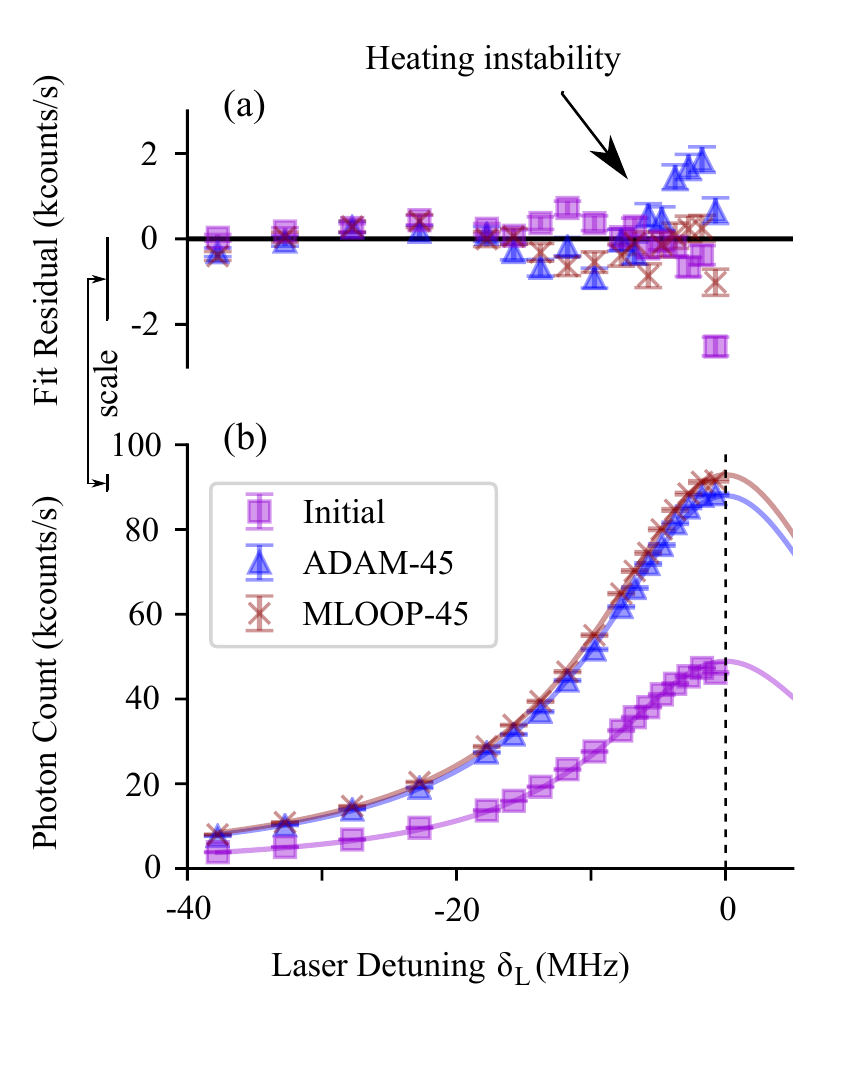}
 \caption{Fluorescence vs laser frequency detuning from the resonance for inital setting and after different optimizations. It can be seen from (a) that the experimental values are very close to the theoretical Lorentzian fit \cite{metcalf_laser_1999, ejtemaee_optimization_2010, berkeland_destabilization_2002}. This shows the heating is low before and after optimization and therefore the change in fluorescence can be used to infer the change in heating. Deviation from the theory near the resonance shown in (b) is a sign of small heating instability.}
\label{res}
\end{figure}

\begin{table}[t]
\begin{tabular}{l|l|l|l|}
\cline{2-4}
                                            & $P_{sat}$ ($\mu W$)  & $\eta$ (\%)     & $\bar{Bkg} (1/s)$ \\ \hline \hline
\multicolumn{1}{|l||}{Initial}               & 4.11(1)
 & 0.971(2) & 1184(34)      \\ \hline
\multicolumn{1}{|l||}{ADAM-45 }  & 1.86(1) & 0.938(1) & 1227(35)     \\ \hline
\multicolumn{1}{|l||}{MLOOP-45 } & 1.86(1) & 0.989(1) & 1291(35)      \\ \hline
 
\end{tabular}
\caption{Saturation power ($P_{sat}$), detection efficiency ($\eta$) and average background count rate ($Bkg$) before and after optimization with different methods. $P_{sat}$ and $\eta$ values are calculated by fitting the theoretical formula for Fluorescence Rate vs Laser Power to the experimental data \cite{metcalf_laser_1999, ejtemaee_optimization_2010, berkeland_destabilization_2002}. A clear drop in saturation power is observed after ADAM and MLOOP optimization of individual electrodes indicating a reduction of the micromotion.}
  \label{SaturationParametersTable}
\end{table}

To test the effectiveness of the protocols, the saturation power, $P_{sat}$, was measured before and after the optimization process. The $P_{sat}$ is the laser power at which the fluorescence rate of a two-level system is half the fluorescence at infinite laser power. We also measured the overall detection efficiency $\eta$, the fraction of emitted photons which resulted in detection events. Table.~\ref{SaturationParametersTable} shows $P_{sat}$ decreased (ion photon absorption was improved) using both ADAM and MLOOP. The detection efficiency was approximately the same for all runs as expected. 

Another test was done by measuring fluorescence vs laser detuning before and after optimization. Figure~\ref{res} shows that the measured values follows the expected Lorentzian profile \cite{metcalf_laser_1999, ejtemaee_optimization_2010, berkeland_destabilization_2002} and associated linewidth before and after optimization. This indicates that the initial micromotion magnitude $\beta$ was sufficiently small for fluorescence to be a good optimization proxy. Clear increase in florescence can be seen after optimizing 44 electrodes individually both with ADAM and MLOOP. The fit residual curve (difference between the experimental values and the theoretical fit) shows that optimizing individual electrodes, resulted in slight increase in heating instability near the resonance.

\subsection{Testing under poor trap conditions}
\label{poor_con}

To test the live performance of the optimization protocol in a non-ideal situation, we deliberately charged the trap by shining 369.5~nm UV laser light onto the chip for 70 minutes. The power of the laser was $200\pm 15\mu W$ and the Gaussian diameter of the focus was $120\pm 10\mu m$. This process ejects electrons due to the photo-electric effect \cite{wang_laser-induced_2011} and produces irregular and potentially unpredictable slow time varying electric fields within the trap. The process charged the trap significantly and made a noticeable reduction to the photon count. The ADAM algorithm was then tested both during charging and after charging was stopped. In both cases an improvement of fluorescence rate was observed.

\begin{figure}
 \includegraphics{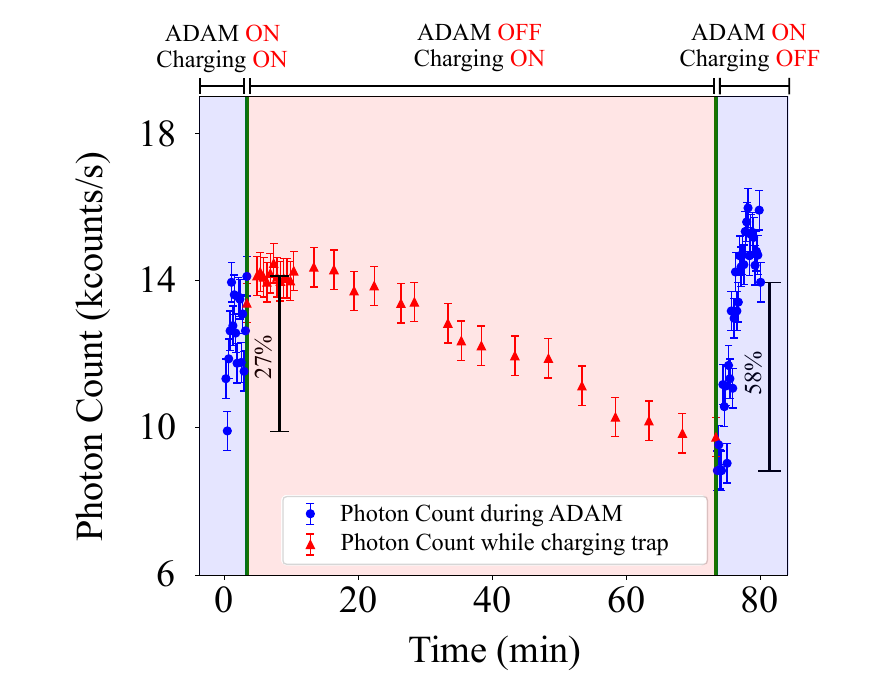}
 \caption{Real time compensation with ADAM of laser charging induced stray electric field. The ion was optimized using ADAM (left blue points) then the photon count was noted whilst charging for 70 minutes (red points) then re-optimized (right blue points). Initial improvement from manually optimized settings was 27\%. The second optimization improved the fluorescence by 58\% from the charged conditions and returned it back to the optimum value of the first optimization within the error.}
 \label{optChargeOpt}
\end{figure}

\begin{figure}
 \includegraphics{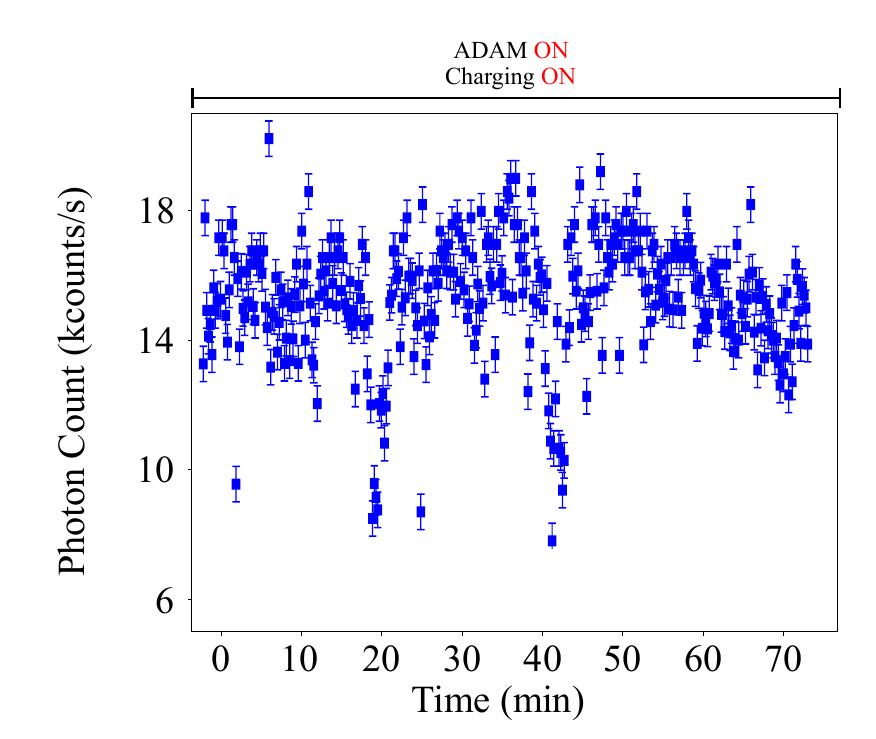}
 \caption{The trap was charged by hitting a UV laser to destabilize the ion and individual electrodes optimized using ADAM simultaneously for 70 minutes. The photon count fluctuates as a result of combination of fluctuations of power of the cooling laser power, algorithm search and charging irregularities. The optimizer keeps the fluorescence at the photon count similar to the case of optimizing after the charging is stopped (third section of Fig.\ref{optChargeOpt}). }
 \label{optWhileCharging}
\end{figure}



The first experiment was performed to test the optimizing process after charging. In this test, starting with the optimal manual setting, ADAM individual electrode optimizer was able to obtain 27\% improvement in the fluorescence rate (blue points on the left side of Fig. \ref{optChargeOpt}). Then charging was induced onto the trap for 70~min and a clear decrease in photon count was seen that went even lower than the initial value (red points in Fig. \ref{optChargeOpt}). At this point charging was stopped and ADAM was run again and fluorescence rate returned back to the previous optimum, within the error, in approximately 12~min. During the second optimization, the fluorescence goes higher than the stable final value for some iterations before the final. This is because of the same effect explained in section \ref{ADAM} that the fluorescence might spike right after a change but go down slightly after stabilizing. Looking at the changes of individual electrodes, shown in Fig.~\ref{ChangeInVoltage}b, we see that the main electrodes adjusted were those around the ion and some throughout the trap. The change in the laser horizontal position was negligible.

Another experiment was done by running ADAM during continuous charging for real-time compensation. Since we induce charging via laser scattering from the trap, the collected photons are both from the ion and the scattered laser and fluctuations in the intensity of scattered light confuses the optimizer. Despite that the optimizer did not lose the ion nor needed to abort the process. Figure~\ref{optWhileCharging} shows that the fluorescence rate, even after a 70-minute charging session, remained near the optimum value. After stopping the charging, the ion remained trapped for more than 8 hours and was intentionally removed from the trap after this time.

\section{Conclusion}
 Comparing the results of the gradient descent method and MLOOP shows that ADAM, the local optimizer, was a better performer in volatile conditions. This is because unlike MLOOP, ADAM does not need any training and can react faster to changes. MLOOP as a global optimizer was able to find settings with higher fluorescence but occasionally optimums were unstable if the defined parameter search range for the settings were large or if the machine learner ran for too long. The main drawback of MLOOP was that it was much slower than ADAM. In both cases photon collection times were the limiting factor.
 
 Optimizing the individual electrodes improved the fluorescence by 78\% for ADAM and 96\% for MLOOP. ADAM showed its reliability in the charging test, returning the ion to the optimal fluorescence in 12 minutes after 70 minutes of purposely charging the trap with a UV laser. Improvement amounts depended greatly on the overall stability of the cooling laser and efficiency of the photon counting system.

To improve the speed of the gradient based optimizer, shorter readout times can be used although this increases variability in the value of the photon count. It is also possible to try local optimizers like SPSA \cite{spall_implementation_1998} that approximate the gradient using only 2 readouts instead of 90. This improves the time needed for each iteration but because using this approximate gradient reduces the improvement in each iteration, the overall speed comparison to reach the same optimum can only be made when both are tried in practice. One possible problem with SPSA is that it cannot handle noise like ADAM and multiple readouts and averaging are needed in noisy conditions. If high amount of averaging is required, the algorithm may become very slow. 

\subsection{Acknowledgement}
This research was financially supported by the Griffith University Research Infrastructure Programme, the Griffith Sciences equipment scheme, and Australian Research Council Linkage (LP180100096) Projects, ML is supported by Australian Research Council Future Fellowship (FT180100055), AZ was supported by Centre for Quantum Dynamics, Griffith University, Summer Scholarship,  JS, and KS were supported by the Australian Government Research Training Program Scholarship.

\FloatBarrier

\bibliography{ref}

\providecommand{\noopsort}[1]{}\providecommand{\singleletter}[1]{#1}%
\begin{thebibliography}{35}%
\makeatletter
\providecommand \@ifxundefined [1]{%
 \@ifx{#1\undefined}
}%
\providecommand \@ifnum [1]{%
 \ifnum #1\expandafter \@firstoftwo
 \else \expandafter \@secondoftwo
 \fi
}%
\providecommand \@ifx [1]{%
 \ifx #1\expandafter \@firstoftwo
 \else \expandafter \@secondoftwo
 \fi
}%
\providecommand \natexlab [1]{#1}%
\providecommand \enquote  [1]{``#1''}%
\providecommand \bibnamefont  [1]{#1}%
\providecommand \bibfnamefont [1]{#1}%
\providecommand \citenamefont [1]{#1}%
\providecommand \href@noop [0]{\@secondoftwo}%
\providecommand \href [0]{\begingroup \@sanitize@url \@href}%
\providecommand \@href[1]{\@@startlink{#1}\@@href}%
\providecommand \@@href[1]{\endgroup#1\@@endlink}%
\providecommand \@sanitize@url [0]{\catcode `\\12\catcode `\$12\catcode
  `\&12\catcode `\#12\catcode `\^12\catcode `\_12\catcode `\%12\relax}%
\providecommand \@@startlink[1]{}%
\providecommand \@@endlink[0]{}%
\providecommand \url  [0]{\begingroup\@sanitize@url \@url }%
\providecommand \@url [1]{\endgroup\@href {#1}{\urlprefix }}%
\providecommand \urlprefix  [0]{URL }%
\providecommand \Eprint [0]{\href }%
\providecommand \doibase [0]{https://doi.org/}%
\providecommand \selectlanguage [0]{\@gobble}%
\providecommand \bibinfo  [0]{\@secondoftwo}%
\providecommand \bibfield  [0]{\@secondoftwo}%
\providecommand \translation [1]{[#1]}%
\providecommand \BibitemOpen [0]{}%
\providecommand \bibitemStop [0]{}%
\providecommand \bibitemNoStop [0]{.\EOS\space}%
\providecommand \EOS [0]{\spacefactor3000\relax}%
\providecommand \BibitemShut  [1]{\csname bibitem#1\endcsname}%
\let\auto@bib@innerbib\@empty
\bibitem [{\citenamefont {Debnath}\ \emph {et~al.}(2016)\citenamefont
  {Debnath}, \citenamefont {Linke}, \citenamefont {Figgatt}, \citenamefont
  {Landsman}, \citenamefont {Wright},\ and\ \citenamefont
  {Monroe}}]{debnath_demonstration_2016}%
  \BibitemOpen
  \bibfield  {author} {\bibinfo {author} {\bibfnamefont {S.}~\bibnamefont
  {Debnath}}, \bibinfo {author} {\bibfnamefont {N.~M.}\ \bibnamefont {Linke}},
  \bibinfo {author} {\bibfnamefont {C.}~\bibnamefont {Figgatt}}, \bibinfo
  {author} {\bibfnamefont {K.~A.}\ \bibnamefont {Landsman}}, \bibinfo {author}
  {\bibfnamefont {K.}~\bibnamefont {Wright}},\ and\ \bibinfo {author}
  {\bibfnamefont {C.}~\bibnamefont {Monroe}},\ }\bibfield  {title} {\bibinfo
  {title} {Demonstration of a small programmable quantum computer with atomic
  qubits},\ }\href {https://doi.org/10.1038/nature18648} {\bibfield  {journal}
  {\bibinfo  {journal} {Nature}\ }\textbf {\bibinfo {volume} {536}},\ \bibinfo
  {pages} {63} (\bibinfo {year} {2016})}\BibitemShut {NoStop}%
\bibitem [{\citenamefont {Lekitsch}\ \emph {et~al.}(2017)\citenamefont
  {Lekitsch}, \citenamefont {Weidt}, \citenamefont {Fowler}, \citenamefont
  {Mølmer}, \citenamefont {Devitt}, \citenamefont {Wunderlich},\ and\
  \citenamefont {Hensinger}}]{lekitsch_blueprint_2017}%
  \BibitemOpen
  \bibfield  {author} {\bibinfo {author} {\bibfnamefont {B.}~\bibnamefont
  {Lekitsch}}, \bibinfo {author} {\bibfnamefont {S.}~\bibnamefont {Weidt}},
  \bibinfo {author} {\bibfnamefont {A.~G.}\ \bibnamefont {Fowler}}, \bibinfo
  {author} {\bibfnamefont {K.}~\bibnamefont {Mølmer}}, \bibinfo {author}
  {\bibfnamefont {S.~J.}\ \bibnamefont {Devitt}}, \bibinfo {author}
  {\bibfnamefont {C.}~\bibnamefont {Wunderlich}},\ and\ \bibinfo {author}
  {\bibfnamefont {W.~K.}\ \bibnamefont {Hensinger}},\ }\bibfield  {title}
  {\bibinfo {title} {Blueprint for a microwave trapped ion quantum computer},\
  }\href {https://doi.org/10.1126/sciadv.1601540} {\bibfield  {journal}
  {\bibinfo  {journal} {Science Advances}\ }\textbf {\bibinfo {volume} {3}},\
  \bibinfo {pages} {e1601540} (\bibinfo {year} {2017})}\BibitemShut {NoStop}%
\bibitem [{\citenamefont {Monroe}\ and\ \citenamefont
  {Kim}(2013)}]{monroe_scaling_2013}%
  \BibitemOpen
  \bibfield  {author} {\bibinfo {author} {\bibfnamefont {C.}~\bibnamefont
  {Monroe}}\ and\ \bibinfo {author} {\bibfnamefont {J.}~\bibnamefont {Kim}},\
  }\bibfield  {title} {\bibinfo {title} {Scaling the {Ion} {Trap} {Quantum}
  {Processor}},\ }\href {https://doi.org/10.1126/science.1231298} {\bibfield
  {journal} {\bibinfo  {journal} {Science}\ }\textbf {\bibinfo {volume}
  {339}},\ \bibinfo {pages} {1164} (\bibinfo {year} {2013})}\BibitemShut
  {NoStop}%
\bibitem [{\citenamefont {Blinov}\ \emph {et~al.}(2004)\citenamefont {Blinov},
  \citenamefont {Leibfried}, \citenamefont {Monroe},\ and\ \citenamefont
  {Wineland}}]{blinov_quantum_2004}%
  \BibitemOpen
  \bibfield  {author} {\bibinfo {author} {\bibfnamefont {B.~B.}\ \bibnamefont
  {Blinov}}, \bibinfo {author} {\bibfnamefont {D.}~\bibnamefont {Leibfried}},
  \bibinfo {author} {\bibfnamefont {C.}~\bibnamefont {Monroe}},\ and\ \bibinfo
  {author} {\bibfnamefont {D.~J.}\ \bibnamefont {Wineland}},\ }\bibfield
  {title} {\bibinfo {title} {Quantum {Computing} with {Trapped} {Ion}
  {Hyperfine} {Qubits}},\ }\href {https://doi.org/10.1007/s11128-004-9417-3}
  {\bibfield  {journal} {\bibinfo  {journal} {Quantum Information Processing}\
  }\textbf {\bibinfo {volume} {3}},\ \bibinfo {pages} {45} (\bibinfo {year}
  {2004})}\BibitemShut {NoStop}%
\bibitem [{\citenamefont {Wineland}\ \emph {et~al.}(1998)\citenamefont
  {Wineland}, \citenamefont {Monroe}, \citenamefont {Itano}, \citenamefont
  {Leibfried}, \citenamefont {King},\ and\ \citenamefont
  {Meekhof}}]{wineland_experimental_1998}%
  \BibitemOpen
  \bibfield  {author} {\bibinfo {author} {\bibfnamefont {D.~J.}\ \bibnamefont
  {Wineland}}, \bibinfo {author} {\bibfnamefont {C.}~\bibnamefont {Monroe}},
  \bibinfo {author} {\bibfnamefont {W.~M.}\ \bibnamefont {Itano}}, \bibinfo
  {author} {\bibfnamefont {D.}~\bibnamefont {Leibfried}}, \bibinfo {author}
  {\bibfnamefont {B.~E.}\ \bibnamefont {King}},\ and\ \bibinfo {author}
  {\bibfnamefont {D.~M.}\ \bibnamefont {Meekhof}},\ }\bibfield  {title}
  {\bibinfo {title} {Experimental {Issues} in {Coherent} {Quantum}-{State}
  {Manipulation} of {Trapped} {Atomic} {Ions}},\ }\href
  {https://doi.org/10.6028/jres.103.019} {\bibfield  {journal} {\bibinfo
  {journal} {Journal of Research of the National Institute of Standards and
  Technology}\ }\textbf {\bibinfo {volume} {103}},\ \bibinfo {pages} {259}
  (\bibinfo {year} {1998})}\BibitemShut {NoStop}%
\bibitem [{\citenamefont {Kielpinski}\ \emph {et~al.}(2002)\citenamefont
  {Kielpinski}, \citenamefont {Monroe},\ and\ \citenamefont
  {Wineland}}]{kielpinski_architecture_2002}%
  \BibitemOpen
  \bibfield  {author} {\bibinfo {author} {\bibfnamefont {D.}~\bibnamefont
  {Kielpinski}}, \bibinfo {author} {\bibfnamefont {C.}~\bibnamefont {Monroe}},\
  and\ \bibinfo {author} {\bibfnamefont {D.~J.}\ \bibnamefont {Wineland}},\
  }\bibfield  {title} {\bibinfo {title} {Architecture for a large-scale
  ion-trap quantum computer},\ }\href {https://doi.org/10.1038/nature00784}
  {\bibfield  {journal} {\bibinfo  {journal} {Nature}\ }\textbf {\bibinfo
  {volume} {417}},\ \bibinfo {pages} {709} (\bibinfo {year}
  {2002})}\BibitemShut {NoStop}%
\bibitem [{\citenamefont {Chiaverini}\ \emph {et~al.}(2005)\citenamefont
  {Chiaverini}, \citenamefont {Blakestad}, \citenamefont {Britton},
  \citenamefont {Jost}, \citenamefont {Langer}, \citenamefont {Leibfried},
  \citenamefont {Ozeri},\ and\ \citenamefont
  {Wineland}}]{chiaverini_surface-electrode_2005}%
  \BibitemOpen
  \bibfield  {author} {\bibinfo {author} {\bibfnamefont {J.}~\bibnamefont
  {Chiaverini}}, \bibinfo {author} {\bibfnamefont {R.~B.}\ \bibnamefont
  {Blakestad}}, \bibinfo {author} {\bibfnamefont {J.}~\bibnamefont {Britton}},
  \bibinfo {author} {\bibfnamefont {J.~D.}\ \bibnamefont {Jost}}, \bibinfo
  {author} {\bibfnamefont {C.}~\bibnamefont {Langer}}, \bibinfo {author}
  {\bibfnamefont {D.}~\bibnamefont {Leibfried}}, \bibinfo {author}
  {\bibfnamefont {R.}~\bibnamefont {Ozeri}},\ and\ \bibinfo {author}
  {\bibfnamefont {D.~J.}\ \bibnamefont {Wineland}},\ }\bibfield  {title}
  {\bibinfo {title} {Surface-electrode architecture for ion-trap quantum
  information processing},\ }\href@noop {} {\bibfield  {journal} {\bibinfo
  {journal} {Quantum Information \& Computation}\ }\textbf {\bibinfo {volume}
  {5}},\ \bibinfo {pages} {419} (\bibinfo {year} {2005})}\BibitemShut {NoStop}%
\bibitem [{\citenamefont {Seidelin}\ \emph {et~al.}(2006)\citenamefont
  {Seidelin}, \citenamefont {Chiaverini}, \citenamefont {Reichle},
  \citenamefont {Bollinger}, \citenamefont {Leibfried}, \citenamefont
  {Britton}, \citenamefont {Wesenberg}, \citenamefont {Blakestad},
  \citenamefont {Epstein}, \citenamefont {Hume}, \citenamefont {Itano},
  \citenamefont {Jost}, \citenamefont {Langer}, \citenamefont {Ozeri},
  \citenamefont {Shiga},\ and\ \citenamefont
  {Wineland}}]{seidelin_microfabricated_2006}%
  \BibitemOpen
  \bibfield  {author} {\bibinfo {author} {\bibfnamefont {S.}~\bibnamefont
  {Seidelin}}, \bibinfo {author} {\bibfnamefont {J.}~\bibnamefont
  {Chiaverini}}, \bibinfo {author} {\bibfnamefont {R.}~\bibnamefont {Reichle}},
  \bibinfo {author} {\bibfnamefont {J.~J.}\ \bibnamefont {Bollinger}}, \bibinfo
  {author} {\bibfnamefont {D.}~\bibnamefont {Leibfried}}, \bibinfo {author}
  {\bibfnamefont {J.}~\bibnamefont {Britton}}, \bibinfo {author} {\bibfnamefont
  {J.~H.}\ \bibnamefont {Wesenberg}}, \bibinfo {author} {\bibfnamefont {R.~B.}\
  \bibnamefont {Blakestad}}, \bibinfo {author} {\bibfnamefont {R.~J.}\
  \bibnamefont {Epstein}}, \bibinfo {author} {\bibfnamefont {D.~B.}\
  \bibnamefont {Hume}}, \bibinfo {author} {\bibfnamefont {W.~M.}\ \bibnamefont
  {Itano}}, \bibinfo {author} {\bibfnamefont {J.~D.}\ \bibnamefont {Jost}},
  \bibinfo {author} {\bibfnamefont {C.}~\bibnamefont {Langer}}, \bibinfo
  {author} {\bibfnamefont {R.}~\bibnamefont {Ozeri}}, \bibinfo {author}
  {\bibfnamefont {N.}~\bibnamefont {Shiga}},\ and\ \bibinfo {author}
  {\bibfnamefont {D.~J.}\ \bibnamefont {Wineland}},\ }\bibfield  {title}
  {\bibinfo {title} {Microfabricated {Surface}-{Electrode} {Ion} {Trap} for
  {Scalable} {Quantum} {Information} {Processing}},\ }\href
  {https://doi.org/10.1103/PhysRevLett.96.253003} {\bibfield  {journal}
  {\bibinfo  {journal} {Physical Review Letters}\ }\textbf {\bibinfo {volume}
  {96}},\ \bibinfo {pages} {253003} (\bibinfo {year} {2006})}\BibitemShut
  {NoStop}%
\bibitem [{\citenamefont {Doret}\ \emph {et~al.}(2012)\citenamefont {Doret},
  \citenamefont {Amini}, \citenamefont {Wright}, \citenamefont {Volin},
  \citenamefont {Killian}, \citenamefont {Ozakin}, \citenamefont {Denison},
  \citenamefont {Hayden}, \citenamefont {Pai}, \citenamefont {Slusher},\ and\
  \citenamefont {Harter}}]{doret_controlling_2012}%
  \BibitemOpen
  \bibfield  {author} {\bibinfo {author} {\bibfnamefont {S.~C.}\ \bibnamefont
  {Doret}}, \bibinfo {author} {\bibfnamefont {J.~M.}\ \bibnamefont {Amini}},
  \bibinfo {author} {\bibfnamefont {K.}~\bibnamefont {Wright}}, \bibinfo
  {author} {\bibfnamefont {C.}~\bibnamefont {Volin}}, \bibinfo {author}
  {\bibfnamefont {T.}~\bibnamefont {Killian}}, \bibinfo {author} {\bibfnamefont
  {A.}~\bibnamefont {Ozakin}}, \bibinfo {author} {\bibfnamefont
  {D.}~\bibnamefont {Denison}}, \bibinfo {author} {\bibfnamefont
  {H.}~\bibnamefont {Hayden}}, \bibinfo {author} {\bibfnamefont {C.-S.}\
  \bibnamefont {Pai}}, \bibinfo {author} {\bibfnamefont {R.~E.}\ \bibnamefont
  {Slusher}},\ and\ \bibinfo {author} {\bibfnamefont {A.~W.}\ \bibnamefont
  {Harter}},\ }\bibfield  {title} {\bibinfo {title} {Controlling trapping
  potentials and stray electric fields in a microfabricated ion trap through
  design and compensation},\ }\href
  {https://doi.org/10.1088/1367-2630/14/7/073012} {\bibfield  {journal}
  {\bibinfo  {journal} {New Journal of Physics}\ }\textbf {\bibinfo {volume}
  {14}},\ \bibinfo {pages} {073012} (\bibinfo {year} {2012})}\BibitemShut
  {NoStop}%
\bibitem [{\citenamefont {Amini}\ \emph {et~al.}(2010)\citenamefont {Amini},
  \citenamefont {Uys}, \citenamefont {Wesenberg}, \citenamefont {Seidelin},
  \citenamefont {Britton}, \citenamefont {Bollinger}, \citenamefont
  {Leibfried}, \citenamefont {Ospelkaus}, \citenamefont {VanDevender},\ and\
  \citenamefont {Wineland}}]{amini_toward_2010}%
  \BibitemOpen
  \bibfield  {author} {\bibinfo {author} {\bibfnamefont {J.~M.}\ \bibnamefont
  {Amini}}, \bibinfo {author} {\bibfnamefont {H.}~\bibnamefont {Uys}}, \bibinfo
  {author} {\bibfnamefont {J.~H.}\ \bibnamefont {Wesenberg}}, \bibinfo {author}
  {\bibfnamefont {S.}~\bibnamefont {Seidelin}}, \bibinfo {author}
  {\bibfnamefont {J.}~\bibnamefont {Britton}}, \bibinfo {author} {\bibfnamefont
  {J.~J.}\ \bibnamefont {Bollinger}}, \bibinfo {author} {\bibfnamefont
  {D.}~\bibnamefont {Leibfried}}, \bibinfo {author} {\bibfnamefont
  {C.}~\bibnamefont {Ospelkaus}}, \bibinfo {author} {\bibfnamefont {A.~P.}\
  \bibnamefont {VanDevender}},\ and\ \bibinfo {author} {\bibfnamefont {D.~J.}\
  \bibnamefont {Wineland}},\ }\bibfield  {title} {\bibinfo {title} {Toward
  scalable ion traps for quantum information processing},\ }\href
  {https://doi.org/10.1088/1367-2630/12/3/033031} {\bibfield  {journal}
  {\bibinfo  {journal} {New Journal of Physics}\ }\textbf {\bibinfo {volume}
  {12}},\ \bibinfo {pages} {033031} (\bibinfo {year} {2010})}\BibitemShut
  {NoStop}%
\bibitem [{\citenamefont {Kielpinski}(2001)}]{kielpinski_entanglement_2001}%
  \BibitemOpen
  \bibfield  {author} {\bibinfo {author} {\bibfnamefont {D.}~\bibnamefont
  {Kielpinski}},\ }\emph {\bibinfo {title} {Entanglement and decoherence in a
  trapped-ion quantum register}},\ \href
  {http://adsabs.harvard.edu/abs/2001PhDT........29K} {Ph.D. thesis} (\bibinfo
  {year} {2001})\BibitemShut {NoStop}%
\bibitem [{\citenamefont {Tanaka}\ \emph {et~al.}(2012)\citenamefont {Tanaka},
  \citenamefont {Masuda}, \citenamefont {Akimoto}, \citenamefont {Koda},
  \citenamefont {Ibaraki},\ and\ \citenamefont
  {Urabe}}]{tanaka_micromotion_2012}%
  \BibitemOpen
  \bibfield  {author} {\bibinfo {author} {\bibfnamefont {U.}~\bibnamefont
  {Tanaka}}, \bibinfo {author} {\bibfnamefont {K.}~\bibnamefont {Masuda}},
  \bibinfo {author} {\bibfnamefont {Y.}~\bibnamefont {Akimoto}}, \bibinfo
  {author} {\bibfnamefont {K.}~\bibnamefont {Koda}}, \bibinfo {author}
  {\bibfnamefont {Y.}~\bibnamefont {Ibaraki}},\ and\ \bibinfo {author}
  {\bibfnamefont {S.}~\bibnamefont {Urabe}},\ }\bibfield  {title} {\bibinfo
  {title} {Micromotion compensation in a surface electrode trap by parametric
  excitation of trapped ions},\ }\href
  {https://doi.org/10.1007/s00340-011-4762-2} {\bibfield  {journal} {\bibinfo
  {journal} {Applied Physics B}\ }\textbf {\bibinfo {volume} {107}},\ \bibinfo
  {pages} {907} (\bibinfo {year} {2012})}\BibitemShut {NoStop}%
\bibitem [{\citenamefont {Allcock}\ \emph {et~al.}(2012)\citenamefont
  {Allcock}, \citenamefont {Harty}, \citenamefont {Janacek}, \citenamefont
  {Linke}, \citenamefont {Ballance}, \citenamefont {Steane}, \citenamefont
  {Lucas}, \citenamefont {Jarecki}, \citenamefont {Habermehl}, \citenamefont
  {Blain}, \citenamefont {Stick},\ and\ \citenamefont
  {Moehring}}]{allcock_heating_2012}%
  \BibitemOpen
  \bibfield  {author} {\bibinfo {author} {\bibfnamefont {D.~T.~C.}\
  \bibnamefont {Allcock}}, \bibinfo {author} {\bibfnamefont {T.~P.}\
  \bibnamefont {Harty}}, \bibinfo {author} {\bibfnamefont {H.~A.}\ \bibnamefont
  {Janacek}}, \bibinfo {author} {\bibfnamefont {N.~M.}\ \bibnamefont {Linke}},
  \bibinfo {author} {\bibfnamefont {C.~J.}\ \bibnamefont {Ballance}}, \bibinfo
  {author} {\bibfnamefont {A.~M.}\ \bibnamefont {Steane}}, \bibinfo {author}
  {\bibfnamefont {D.~M.}\ \bibnamefont {Lucas}}, \bibinfo {author}
  {\bibfnamefont {R.~L.}\ \bibnamefont {Jarecki}}, \bibinfo {author}
  {\bibfnamefont {S.~D.}\ \bibnamefont {Habermehl}}, \bibinfo {author}
  {\bibfnamefont {M.~G.}\ \bibnamefont {Blain}}, \bibinfo {author}
  {\bibfnamefont {D.}~\bibnamefont {Stick}},\ and\ \bibinfo {author}
  {\bibfnamefont {D.~L.}\ \bibnamefont {Moehring}},\ }\bibfield  {title}
  {\bibinfo {title} {Heating rate and electrode charging measurements in a
  scalable, microfabricated, surface-electrode ion trap},\ }\href
  {https://doi.org/10.1007/s00340-011-4788-5} {\bibfield  {journal} {\bibinfo
  {journal} {Applied Physics B}\ }\textbf {\bibinfo {volume} {107}},\ \bibinfo
  {pages} {913} (\bibinfo {year} {2012})}\BibitemShut {NoStop}%
\bibitem [{\citenamefont {Narayanan}\ \emph {et~al.}(2011)\citenamefont
  {Narayanan}, \citenamefont {Daniilidis}, \citenamefont {Möller},
  \citenamefont {Clark}, \citenamefont {Ziesel}, \citenamefont {Singer},
  \citenamefont {Schmidt-Kaler},\ and\ \citenamefont
  {Häffner}}]{narayanan_electric_2011}%
  \BibitemOpen
  \bibfield  {author} {\bibinfo {author} {\bibfnamefont {S.}~\bibnamefont
  {Narayanan}}, \bibinfo {author} {\bibfnamefont {N.}~\bibnamefont
  {Daniilidis}}, \bibinfo {author} {\bibfnamefont {S.~A.}\ \bibnamefont
  {Möller}}, \bibinfo {author} {\bibfnamefont {R.}~\bibnamefont {Clark}},
  \bibinfo {author} {\bibfnamefont {F.}~\bibnamefont {Ziesel}}, \bibinfo
  {author} {\bibfnamefont {K.}~\bibnamefont {Singer}}, \bibinfo {author}
  {\bibfnamefont {F.}~\bibnamefont {Schmidt-Kaler}},\ and\ \bibinfo {author}
  {\bibfnamefont {H.}~\bibnamefont {Häffner}},\ }\bibfield  {title} {\bibinfo
  {title} {Electric field compensation and sensing with a single ion in a
  planar trap},\ }\href {https://doi.org/10.1063/1.3665647} {\bibfield
  {journal} {\bibinfo  {journal} {Journal of Applied Physics}\ }\textbf
  {\bibinfo {volume} {110}},\ \bibinfo {pages} {114909} (\bibinfo {year}
  {2011})}\BibitemShut {NoStop}%
\bibitem [{\citenamefont {Berkeland}\ \emph {et~al.}(1998)\citenamefont
  {Berkeland}, \citenamefont {Miller}, \citenamefont {Bergquist}, \citenamefont
  {Itano},\ and\ \citenamefont {Wineland}}]{berkeland_minimization_1998}%
  \BibitemOpen
  \bibfield  {author} {\bibinfo {author} {\bibfnamefont {D.~J.}\ \bibnamefont
  {Berkeland}}, \bibinfo {author} {\bibfnamefont {J.~D.}\ \bibnamefont
  {Miller}}, \bibinfo {author} {\bibfnamefont {J.~C.}\ \bibnamefont
  {Bergquist}}, \bibinfo {author} {\bibfnamefont {W.~M.}\ \bibnamefont
  {Itano}},\ and\ \bibinfo {author} {\bibfnamefont {D.~J.}\ \bibnamefont
  {Wineland}},\ }\bibfield  {title} {\bibinfo {title} {Minimization of ion
  micromotion in a {Paul} trap},\ }\href {https://doi.org/10.1063/1.367318}
  {\bibfield  {journal} {\bibinfo  {journal} {Journal of Applied Physics}\
  }\textbf {\bibinfo {volume} {83}},\ \bibinfo {pages} {5025} (\bibinfo {year}
  {1998})}\BibitemShut {NoStop}%
\bibitem [{\citenamefont {Noek}\ \emph {et~al.}(2013)\citenamefont {Noek},
  \citenamefont {Kim}, \citenamefont {Mount}, \citenamefont {Baek},
  \citenamefont {Maunz},\ and\ \citenamefont {Kim}}]{noek_trapping_2013}%
  \BibitemOpen
  \bibfield  {author} {\bibinfo {author} {\bibfnamefont {R.}~\bibnamefont
  {Noek}}, \bibinfo {author} {\bibfnamefont {T.}~\bibnamefont {Kim}}, \bibinfo
  {author} {\bibfnamefont {E.}~\bibnamefont {Mount}}, \bibinfo {author}
  {\bibfnamefont {S.-Y.}\ \bibnamefont {Baek}}, \bibinfo {author}
  {\bibfnamefont {P.}~\bibnamefont {Maunz}},\ and\ \bibinfo {author}
  {\bibfnamefont {J.}~\bibnamefont {Kim}},\ }\bibfield  {title} {\bibinfo
  {title} {Trapping and cooling of {174Yb}+ ions in a microfabricated surface
  trap},\ }\href {https://doi.org/10.3938/jkps.63.907} {\bibfield  {journal}
  {\bibinfo  {journal} {Journal of the Korean Physical Society}\ }\textbf
  {\bibinfo {volume} {63}},\ \bibinfo {pages} {907} (\bibinfo {year}
  {2013})}\BibitemShut {NoStop}%
\bibitem [{\citenamefont {Allcock}\ \emph {et~al.}(2010)\citenamefont
  {Allcock}, \citenamefont {Sherman}, \citenamefont {Stacey}, \citenamefont
  {Burrell}, \citenamefont {Curtis}, \citenamefont {Imreh}, \citenamefont
  {Linke}, \citenamefont {Szwer}, \citenamefont {Webster}, \citenamefont
  {Steane},\ and\ \citenamefont {Lucas}}]{allcock_implementation_2010}%
  \BibitemOpen
  \bibfield  {author} {\bibinfo {author} {\bibfnamefont {D.~T.~C.}\
  \bibnamefont {Allcock}}, \bibinfo {author} {\bibfnamefont {J.~A.}\
  \bibnamefont {Sherman}}, \bibinfo {author} {\bibfnamefont {D.~N.}\
  \bibnamefont {Stacey}}, \bibinfo {author} {\bibfnamefont {A.~H.}\
  \bibnamefont {Burrell}}, \bibinfo {author} {\bibfnamefont {M.~J.}\
  \bibnamefont {Curtis}}, \bibinfo {author} {\bibfnamefont {G.}~\bibnamefont
  {Imreh}}, \bibinfo {author} {\bibfnamefont {N.~M.}\ \bibnamefont {Linke}},
  \bibinfo {author} {\bibfnamefont {D.~J.}\ \bibnamefont {Szwer}}, \bibinfo
  {author} {\bibfnamefont {S.~C.}\ \bibnamefont {Webster}}, \bibinfo {author}
  {\bibfnamefont {A.~M.}\ \bibnamefont {Steane}},\ and\ \bibinfo {author}
  {\bibfnamefont {D.~M.}\ \bibnamefont {Lucas}},\ }\bibfield  {title} {\bibinfo
  {title} {Implementation of a symmetric surface-electrode ion trap with field
  compensation using a modulated {Raman} effect},\ }\href
  {https://doi.org/10.1088/1367-2630/12/5/053026} {\bibfield  {journal}
  {\bibinfo  {journal} {New Journal of Physics}\ }\textbf {\bibinfo {volume}
  {12}},\ \bibinfo {pages} {053026} (\bibinfo {year} {2010})}\BibitemShut
  {NoStop}%
\bibitem [{\citenamefont {Harlander}\ \emph {et~al.}(2010)\citenamefont
  {Harlander}, \citenamefont {Brownnutt}, \citenamefont {Hänsel},\ and\
  \citenamefont {Blatt}}]{harlander_trapped-ion_2010}%
  \BibitemOpen
  \bibfield  {author} {\bibinfo {author} {\bibfnamefont {M.}~\bibnamefont
  {Harlander}}, \bibinfo {author} {\bibfnamefont {M.}~\bibnamefont
  {Brownnutt}}, \bibinfo {author} {\bibfnamefont {W.}~\bibnamefont {Hänsel}},\
  and\ \bibinfo {author} {\bibfnamefont {R.}~\bibnamefont {Blatt}},\ }\bibfield
   {title} {\bibinfo {title} {Trapped-ion probing of light-induced charging
  effects on dielectrics},\ }\href
  {https://doi.org/10.1088/1367-2630/12/9/093035} {\bibfield  {journal}
  {\bibinfo  {journal} {New Journal of Physics}\ }\textbf {\bibinfo {volume}
  {12}},\ \bibinfo {pages} {093035} (\bibinfo {year} {2010})},\ \bibinfo {note}
  {arXiv: 1004.4842}\BibitemShut {NoStop}%
\bibitem [{\citenamefont {Seif}\ \emph {et~al.}(2018)\citenamefont {Seif},
  \citenamefont {Landsman}, \citenamefont {Linke}, \citenamefont {Figgatt},
  \citenamefont {Monroe},\ and\ \citenamefont {Hafezi}}]{seif_machine_2018}%
  \BibitemOpen
  \bibfield  {author} {\bibinfo {author} {\bibfnamefont {A.}~\bibnamefont
  {Seif}}, \bibinfo {author} {\bibfnamefont {K.~A.}\ \bibnamefont {Landsman}},
  \bibinfo {author} {\bibfnamefont {N.~M.}\ \bibnamefont {Linke}}, \bibinfo
  {author} {\bibfnamefont {C.}~\bibnamefont {Figgatt}}, \bibinfo {author}
  {\bibfnamefont {C.}~\bibnamefont {Monroe}},\ and\ \bibinfo {author}
  {\bibfnamefont {M.}~\bibnamefont {Hafezi}},\ }\bibfield  {title} {\bibinfo
  {title} {Machine learning assisted readout of trapped-ion qubits},\ }\href
  {https://doi.org/10.1088/1361-6455/aad62b} {\bibfield  {journal} {\bibinfo
  {journal} {Journal of Physics B: Atomic, Molecular and Optical Physics}\
  }\textbf {\bibinfo {volume} {51}},\ \bibinfo {pages} {174006} (\bibinfo
  {year} {2018})}\BibitemShut {NoStop}%
\bibitem [{\citenamefont {Wigley}\ \emph {et~al.}(2016)\citenamefont {Wigley},
  \citenamefont {Everitt}, \citenamefont {van~den Hengel}, \citenamefont
  {Bastian}, \citenamefont {Sooriyabandara}, \citenamefont {McDonald},
  \citenamefont {Hardman}, \citenamefont {Quinlivan}, \citenamefont {Manju},
  \citenamefont {Kuhn}, \citenamefont {Petersen}, \citenamefont {Luiten},
  \citenamefont {Hope}, \citenamefont {Robins},\ and\ \citenamefont
  {Hush}}]{wigley_fast_2016}%
  \BibitemOpen
  \bibfield  {author} {\bibinfo {author} {\bibfnamefont {P.~B.}\ \bibnamefont
  {Wigley}}, \bibinfo {author} {\bibfnamefont {P.~J.}\ \bibnamefont {Everitt}},
  \bibinfo {author} {\bibfnamefont {A.}~\bibnamefont {van~den Hengel}},
  \bibinfo {author} {\bibfnamefont {J.~W.}\ \bibnamefont {Bastian}}, \bibinfo
  {author} {\bibfnamefont {M.~A.}\ \bibnamefont {Sooriyabandara}}, \bibinfo
  {author} {\bibfnamefont {G.~D.}\ \bibnamefont {McDonald}}, \bibinfo {author}
  {\bibfnamefont {K.~S.}\ \bibnamefont {Hardman}}, \bibinfo {author}
  {\bibfnamefont {C.~D.}\ \bibnamefont {Quinlivan}}, \bibinfo {author}
  {\bibfnamefont {P.}~\bibnamefont {Manju}}, \bibinfo {author} {\bibfnamefont
  {C.~C.~N.}\ \bibnamefont {Kuhn}}, \bibinfo {author} {\bibfnamefont {I.~R.}\
  \bibnamefont {Petersen}}, \bibinfo {author} {\bibfnamefont {A.~N.}\
  \bibnamefont {Luiten}}, \bibinfo {author} {\bibfnamefont {J.~J.}\
  \bibnamefont {Hope}}, \bibinfo {author} {\bibfnamefont {N.~P.}\ \bibnamefont
  {Robins}},\ and\ \bibinfo {author} {\bibfnamefont {M.~R.}\ \bibnamefont
  {Hush}},\ }\bibfield  {title} {\bibinfo {title} {Fast machine-learning online
  optimization of ultra-cold-atom experiments},\ }\href
  {https://doi.org/10.1038/srep25890} {\bibfield  {journal} {\bibinfo
  {journal} {Scientific Reports}\ }\textbf {\bibinfo {volume} {6}},\ \bibinfo
  {pages} {25890} (\bibinfo {year} {2016})}\BibitemShut {NoStop}%
\bibitem [{\citenamefont {Tranter}\ \emph {et~al.}(2018)\citenamefont
  {Tranter}, \citenamefont {Slatyer}, \citenamefont {Hush}, \citenamefont
  {Leung}, \citenamefont {Everett}, \citenamefont {Paul}, \citenamefont
  {Vernaz-Gris}, \citenamefont {Lam}, \citenamefont {Buchler},\ and\
  \citenamefont {Campbell}}]{tranter_multiparameter_2018}%
  \BibitemOpen
  \bibfield  {author} {\bibinfo {author} {\bibfnamefont {A.~D.}\ \bibnamefont
  {Tranter}}, \bibinfo {author} {\bibfnamefont {H.~J.}\ \bibnamefont
  {Slatyer}}, \bibinfo {author} {\bibfnamefont {M.~R.}\ \bibnamefont {Hush}},
  \bibinfo {author} {\bibfnamefont {A.~C.}\ \bibnamefont {Leung}}, \bibinfo
  {author} {\bibfnamefont {J.~L.}\ \bibnamefont {Everett}}, \bibinfo {author}
  {\bibfnamefont {K.~V.}\ \bibnamefont {Paul}}, \bibinfo {author}
  {\bibfnamefont {P.}~\bibnamefont {Vernaz-Gris}}, \bibinfo {author}
  {\bibfnamefont {P.~K.}\ \bibnamefont {Lam}}, \bibinfo {author} {\bibfnamefont
  {B.~C.}\ \bibnamefont {Buchler}},\ and\ \bibinfo {author} {\bibfnamefont
  {G.~T.}\ \bibnamefont {Campbell}},\ }\bibfield  {title} {\bibinfo {title}
  {Multiparameter optimisation of a magneto-optical trap using deep learning},\
  }\href {https://doi.org/10.1038/s41467-018-06847-1} {\bibfield  {journal}
  {\bibinfo  {journal} {Nature Communications}\ }\textbf {\bibinfo {volume}
  {9}},\ \bibinfo {pages} {4360} (\bibinfo {year} {2018})}\BibitemShut
  {NoStop}%
\bibitem [{\citenamefont {Ghadimi}\ \emph {et~al.}(2017)\citenamefont
  {Ghadimi}, \citenamefont {Blūms}, \citenamefont {Norton}, \citenamefont
  {Fisher}, \citenamefont {Connell}, \citenamefont {Amini}, \citenamefont
  {Volin}, \citenamefont {Hayden}, \citenamefont {Pai}, \citenamefont
  {Kielpinski}, \citenamefont {Lobino},\ and\ \citenamefont
  {Streed}}]{ghadimi_scalable_2017}%
  \BibitemOpen
  \bibfield  {author} {\bibinfo {author} {\bibfnamefont {M.}~\bibnamefont
  {Ghadimi}}, \bibinfo {author} {\bibfnamefont {V.}~\bibnamefont {Blūms}},
  \bibinfo {author} {\bibfnamefont {B.~G.}\ \bibnamefont {Norton}}, \bibinfo
  {author} {\bibfnamefont {P.~M.}\ \bibnamefont {Fisher}}, \bibinfo {author}
  {\bibfnamefont {S.~C.}\ \bibnamefont {Connell}}, \bibinfo {author}
  {\bibfnamefont {J.~M.}\ \bibnamefont {Amini}}, \bibinfo {author}
  {\bibfnamefont {C.}~\bibnamefont {Volin}}, \bibinfo {author} {\bibfnamefont
  {H.}~\bibnamefont {Hayden}}, \bibinfo {author} {\bibfnamefont {C.-S.}\
  \bibnamefont {Pai}}, \bibinfo {author} {\bibfnamefont {D.}~\bibnamefont
  {Kielpinski}}, \bibinfo {author} {\bibfnamefont {M.}~\bibnamefont {Lobino}},\
  and\ \bibinfo {author} {\bibfnamefont {E.~W.}\ \bibnamefont {Streed}},\
  }\bibfield  {title} {\bibinfo {title} {Scalable ion–photon quantum
  interface based on integrated diffractive mirrors},\ }\href
  {https://doi.org/10.1038/s41534-017-0006-6} {\bibfield  {journal} {\bibinfo
  {journal} {npj Quantum Information}\ }\textbf {\bibinfo {volume} {3}},\
  \bibinfo {pages} {1} (\bibinfo {year} {2017})}\BibitemShut {NoStop}%
\bibitem [{\citenamefont {Kingma}\ and\ \citenamefont
  {Ba}(2017)}]{kingma_ADAM_2017}%
  \BibitemOpen
  \bibfield  {author} {\bibinfo {author} {\bibfnamefont {D.~P.}\ \bibnamefont
  {Kingma}}\ and\ \bibinfo {author} {\bibfnamefont {J.}~\bibnamefont {Ba}},\
  }\bibfield  {title} {\bibinfo {title} {Adam: {A} {Method} for {Stochastic}
  {Optimization}},\ }\href {http://arxiv.org/abs/1412.6980} {\bibfield
  {journal} {\bibinfo  {journal} {arXiv:1412.6980 [cs]}\ } (\bibinfo {year}
  {2017})},\ \bibinfo {note} {arXiv: 1412.6980}\BibitemShut {NoStop}%
\bibitem [{\citenamefont {Shappert}\ \emph {et~al.}(2013)\citenamefont
  {Shappert}, \citenamefont {Merrill}, \citenamefont {Brown}, \citenamefont
  {Amini}, \citenamefont {Volin}, \citenamefont {Doret}, \citenamefont
  {Hayden}, \citenamefont {Pai}, \citenamefont {Brown},\ and\ \citenamefont
  {Harter}}]{shappert_spatially_2013}%
  \BibitemOpen
  \bibfield  {author} {\bibinfo {author} {\bibfnamefont {C.~M.}\ \bibnamefont
  {Shappert}}, \bibinfo {author} {\bibfnamefont {J.~T.}\ \bibnamefont
  {Merrill}}, \bibinfo {author} {\bibfnamefont {K.~R.}\ \bibnamefont {Brown}},
  \bibinfo {author} {\bibfnamefont {J.~M.}\ \bibnamefont {Amini}}, \bibinfo
  {author} {\bibfnamefont {C.}~\bibnamefont {Volin}}, \bibinfo {author}
  {\bibfnamefont {S.~C.}\ \bibnamefont {Doret}}, \bibinfo {author}
  {\bibfnamefont {H.}~\bibnamefont {Hayden}}, \bibinfo {author} {\bibfnamefont
  {C.-S.}\ \bibnamefont {Pai}}, \bibinfo {author} {\bibfnamefont {K.~R.}\
  \bibnamefont {Brown}},\ and\ \bibinfo {author} {\bibfnamefont {A.~W.}\
  \bibnamefont {Harter}},\ }\bibfield  {title} {\bibinfo {title} {Spatially
  uniform single-qubit gate operations with near-field microwaves and composite
  pulse compensation},\ }\href {https://doi.org/10.1088/1367-2630/15/8/083053}
  {\bibfield  {journal} {\bibinfo  {journal} {New Journal of Physics}\ }\textbf
  {\bibinfo {volume} {15}},\ \bibinfo {pages} {083053} (\bibinfo {year}
  {2013})}\BibitemShut {NoStop}%
\bibitem [{\citenamefont {Blūms}\ \emph {et~al.}(2020)\citenamefont {Blūms},
  \citenamefont {Scarabel}, \citenamefont {Shimizu}, \citenamefont {Ghadimi},
  \citenamefont {Connell}, \citenamefont {Händel}, \citenamefont {Norton},
  \citenamefont {Bridge}, \citenamefont {Kielpinski}, \citenamefont {Lobino},\
  and\ \citenamefont {Streed}}]{blums_laser_2020}%
  \BibitemOpen
  \bibfield  {author} {\bibinfo {author} {\bibfnamefont {V.}~\bibnamefont
  {Blūms}}, \bibinfo {author} {\bibfnamefont {J.}~\bibnamefont {Scarabel}},
  \bibinfo {author} {\bibfnamefont {K.}~\bibnamefont {Shimizu}}, \bibinfo
  {author} {\bibfnamefont {M.}~\bibnamefont {Ghadimi}}, \bibinfo {author}
  {\bibfnamefont {S.~C.}\ \bibnamefont {Connell}}, \bibinfo {author}
  {\bibfnamefont {S.}~\bibnamefont {Händel}}, \bibinfo {author} {\bibfnamefont
  {B.~G.}\ \bibnamefont {Norton}}, \bibinfo {author} {\bibfnamefont {E.~M.}\
  \bibnamefont {Bridge}}, \bibinfo {author} {\bibfnamefont {D.}~\bibnamefont
  {Kielpinski}}, \bibinfo {author} {\bibfnamefont {M.}~\bibnamefont {Lobino}},\
  and\ \bibinfo {author} {\bibfnamefont {E.~W.}\ \bibnamefont {Streed}},\
  }\bibfield  {title} {\bibinfo {title} {Laser stabilization to neutral {Yb} in
  a discharge with polarization-enhanced frequency modulation spectroscopy},\
  }\href {https://doi.org/10.1063/5.0019252} {\bibfield  {journal} {\bibinfo
  {journal} {Review of Scientific Instruments}\ }\textbf {\bibinfo {volume}
  {91}},\ \bibinfo {pages} {123002} (\bibinfo {year} {2020})}\BibitemShut
  {NoStop}%
\bibitem [{\citenamefont {Norton}\ \emph {et~al.}(2011)\citenamefont {Norton},
  \citenamefont {Streed}, \citenamefont {Petrasiunas}, \citenamefont {Jechow},\
  and\ \citenamefont {Kielpinski}}]{norton_millikelvin_2011}%
  \BibitemOpen
  \bibfield  {author} {\bibinfo {author} {\bibfnamefont {B.~G.}\ \bibnamefont
  {Norton}}, \bibinfo {author} {\bibfnamefont {E.~W.}\ \bibnamefont {Streed}},
  \bibinfo {author} {\bibfnamefont {M.~J.}\ \bibnamefont {Petrasiunas}},
  \bibinfo {author} {\bibfnamefont {A.}~\bibnamefont {Jechow}},\ and\ \bibinfo
  {author} {\bibfnamefont {D.}~\bibnamefont {Kielpinski}},\ }\bibfield  {title}
  {\bibinfo {title} {Millikelvin spatial thermometry of trapped ions},\ }\href
  {https://doi.org/10.1088/1367-2630/13/11/113022} {\bibfield  {journal}
  {\bibinfo  {journal} {New Journal of Physics}\ }\textbf {\bibinfo {volume}
  {13}},\ \bibinfo {pages} {113022} (\bibinfo {year} {2011})}\BibitemShut
  {NoStop}%
\bibitem [{\citenamefont {Budker}\ \emph {et~al.}(2004)\citenamefont {Budker},
  \citenamefont {Kimball}, \citenamefont {Kimball},\ and\ \citenamefont
  {DeMille}}]{budker_atomic_2004}%
  \BibitemOpen
  \bibfield  {author} {\bibinfo {author} {\bibfnamefont {D.}~\bibnamefont
  {Budker}}, \bibinfo {author} {\bibfnamefont {D.}~\bibnamefont {Kimball}},
  \bibinfo {author} {\bibfnamefont {D.~F.}\ \bibnamefont {Kimball}},\ and\
  \bibinfo {author} {\bibfnamefont {D.~P.}\ \bibnamefont {DeMille}},\
  }\href@noop {} {\emph {\bibinfo {title} {Atomic {Physics}: {An} {Exploration}
  {Through} {Problems} and {Solutions}}}}\ (\bibinfo  {publisher} {Oxford
  University Press},\ \bibinfo {year} {2004})\BibitemShut {NoStop}%
\bibitem [{\citenamefont {Zappacosta}(2021)}]{github_Zapp}%
  \BibitemOpen
  \bibfield  {author} {\bibinfo {author} {\bibfnamefont {A.}~\bibnamefont
  {Zappacosta}},\ }\href@noop {} {\bibinfo {title} {Auto ion compensation}},\
  \bibinfo {howpublished} {\url{github.com/alexzappa11/Auto_Ion_Compensation}}
  (\bibinfo {year} {2021})\BibitemShut {NoStop}%
\bibitem [{\citenamefont {Ghadimi}\ \emph {et~al.}(2020)\citenamefont
  {Ghadimi}, \citenamefont {Bridge}, \citenamefont {Scarabel}, \citenamefont
  {Connell}, \citenamefont {Shimizu}, \citenamefont {Streed}, \citenamefont
  {Streed}, \citenamefont {Lobino},\ and\ \citenamefont
  {Lobino}}]{ghadimi_multichannel_2020}%
  \BibitemOpen
  \bibfield  {author} {\bibinfo {author} {\bibfnamefont {M.}~\bibnamefont
  {Ghadimi}}, \bibinfo {author} {\bibfnamefont {E.~M.}\ \bibnamefont {Bridge}},
  \bibinfo {author} {\bibfnamefont {J.}~\bibnamefont {Scarabel}}, \bibinfo
  {author} {\bibfnamefont {S.}~\bibnamefont {Connell}}, \bibinfo {author}
  {\bibfnamefont {K.}~\bibnamefont {Shimizu}}, \bibinfo {author} {\bibfnamefont
  {E.}~\bibnamefont {Streed}}, \bibinfo {author} {\bibfnamefont
  {E.}~\bibnamefont {Streed}}, \bibinfo {author} {\bibfnamefont
  {M.}~\bibnamefont {Lobino}},\ and\ \bibinfo {author} {\bibfnamefont
  {M.}~\bibnamefont {Lobino}},\ }\bibfield  {title} {\bibinfo {title}
  {Multichannel optomechanical switch and locking system for wavemeters},\
  }\href {https://doi.org/10.1364/AO.390881} {\bibfield  {journal} {\bibinfo
  {journal} {Applied Optics}\ }\textbf {\bibinfo {volume} {59}},\ \bibinfo
  {pages} {5136} (\bibinfo {year} {2020})}\BibitemShut {NoStop}%
\bibitem [{\citenamefont {Price}\ \emph {et~al.}(2005)\citenamefont {Price},
  \citenamefont {Storn},\ and\ \citenamefont
  {Lampinen}}]{price_differential_2005}%
  \BibitemOpen
  \bibfield  {author} {\bibinfo {author} {\bibfnamefont {K.}~\bibnamefont
  {Price}}, \bibinfo {author} {\bibfnamefont {R.~M.}\ \bibnamefont {Storn}},\
  and\ \bibinfo {author} {\bibfnamefont {J.~A.}\ \bibnamefont {Lampinen}},\
  }\href {https://doi.org/10.1007/3-540-31306-0} {\emph {\bibinfo {title}
  {Differential {Evolution}: {A} {Practical} {Approach} to {Global}
  {Optimization}}}},\ Natural {Computing} {Series}\ (\bibinfo  {publisher}
  {Springer-Verlag},\ \bibinfo {address} {Berlin Heidelberg},\ \bibinfo {year}
  {2005})\BibitemShut {NoStop}%
\bibitem [{\citenamefont {Metcalf}\ \emph {et~al.}(1999)\citenamefont
  {Metcalf}, \citenamefont {Straten},\ and\ \citenamefont
  {Straten}}]{metcalf_laser_1999}%
  \BibitemOpen
  \bibfield  {author} {\bibinfo {author} {\bibfnamefont {H.~J.}\ \bibnamefont
  {Metcalf}}, \bibinfo {author} {\bibfnamefont {P.~v.~d.}\ \bibnamefont
  {Straten}},\ and\ \bibinfo {author} {\bibfnamefont {P.~v.~d.}\ \bibnamefont
  {Straten}},\ }\href {https://doi.org/10.1007/978-1-4612-1470-0} {\emph
  {\bibinfo {title} {Laser {Cooling} and {Trapping}}}},\ Graduate {Texts} in
  {Contemporary} {Physics}\ (\bibinfo  {publisher} {Springer-Verlag},\ \bibinfo
  {address} {New York},\ \bibinfo {year} {1999})\BibitemShut {NoStop}%
\bibitem [{\citenamefont {Ejtemaee}\ \emph {et~al.}(2010)\citenamefont
  {Ejtemaee}, \citenamefont {Thomas},\ and\ \citenamefont
  {Haljan}}]{ejtemaee_optimization_2010}%
  \BibitemOpen
  \bibfield  {author} {\bibinfo {author} {\bibfnamefont {S.}~\bibnamefont
  {Ejtemaee}}, \bibinfo {author} {\bibfnamefont {R.}~\bibnamefont {Thomas}},\
  and\ \bibinfo {author} {\bibfnamefont {P.~C.}\ \bibnamefont {Haljan}},\
  }\bibfield  {title} {\bibinfo {title} {Optimization of yb171+ fluorescence
  and hyperfine-qubit detection},\ }\href
  {https://doi.org/10.1103/PhysRevA.82.063419} {\bibfield  {journal} {\bibinfo
  {journal} {Physical Review A}\ }\textbf {\bibinfo {volume} {82}},\ \bibinfo
  {pages} {063419} (\bibinfo {year} {2010})}\BibitemShut {NoStop}%
\bibitem [{\citenamefont {Berkeland}\ and\ \citenamefont
  {Boshier}(2002)}]{berkeland_destabilization_2002}%
  \BibitemOpen
  \bibfield  {author} {\bibinfo {author} {\bibfnamefont {D.~J.}\ \bibnamefont
  {Berkeland}}\ and\ \bibinfo {author} {\bibfnamefont {M.~G.}\ \bibnamefont
  {Boshier}},\ }\bibfield  {title} {\bibinfo {title} {Destabilization of dark
  states and optical spectroscopy in {Zeeman}-degenerate atomic systems},\
  }\href {https://doi.org/10.1103/PhysRevA.65.033413} {\bibfield  {journal}
  {\bibinfo  {journal} {Physical Review A}\ }\textbf {\bibinfo {volume} {65}},\
  \bibinfo {pages} {033413} (\bibinfo {year} {2002})}\BibitemShut {NoStop}%
\bibitem [{\citenamefont {Wang}\ \emph {et~al.}(2011)\citenamefont {Wang},
  \citenamefont {Low}, \citenamefont {Lachenmyer}, \citenamefont {Ge},
  \citenamefont {Herskind},\ and\ \citenamefont
  {Chuang}}]{wang_laser-induced_2011}%
  \BibitemOpen
  \bibfield  {author} {\bibinfo {author} {\bibfnamefont {S.~X.}\ \bibnamefont
  {Wang}}, \bibinfo {author} {\bibfnamefont {G.~H.}\ \bibnamefont {Low}},
  \bibinfo {author} {\bibfnamefont {N.~S.}\ \bibnamefont {Lachenmyer}},
  \bibinfo {author} {\bibfnamefont {Y.}~\bibnamefont {Ge}}, \bibinfo {author}
  {\bibfnamefont {P.~F.}\ \bibnamefont {Herskind}},\ and\ \bibinfo {author}
  {\bibfnamefont {I.~L.}\ \bibnamefont {Chuang}},\ }\bibfield  {title}
  {\bibinfo {title} {Laser-induced charging of microfabricated ion traps},\
  }\href {https://doi.org/10.1063/1.3662118} {\bibfield  {journal} {\bibinfo
  {journal} {Journal of Applied Physics}\ }\textbf {\bibinfo {volume} {110}},\
  \bibinfo {pages} {104901} (\bibinfo {year} {2011})},\ \bibinfo {note} {arXiv:
  1108.0092}\BibitemShut {NoStop}%
\bibitem [{\citenamefont {Spall}(1998)}]{spall_implementation_1998}%
  \BibitemOpen
  \bibfield  {author} {\bibinfo {author} {\bibfnamefont {J.~C.}\ \bibnamefont
  {Spall}},\ }\bibfield  {title} {\bibinfo {title} {Implementation of the
  simultaneous perturbation algorithm for stochastic optimization},\ }\href
  {https://doi.org/10.1109/7.705889} {\bibfield  {journal} {\bibinfo  {journal}
  {IEEE Transactions on Aerospace and Electronic Systems}\ }\textbf {\bibinfo
  {volume} {34}},\ \bibinfo {pages} {817} (\bibinfo {year} {1998})}\BibitemShut
  {NoStop}%
\end{thebibliography}%
\end{document}